\documentclass[sigconf,noacm]{acmart}

 \newcommand*\numcircledmod[1]{\raisebox{.5pt}{\textcircled{\raisebox{-.9pt} {#1}}}}

\setcopyright{none}

\usepackage{color,soul}
\usepackage{subfig}
\usepackage{algorithm}
\usepackage[noend]{algpseudocode}
\usepackage{fancyvrb}

\begin{document}
\title{Grid-AR: A Grid--based Booster for Learned Cardinality Estimation and Range Joins}

\cfoot{\thepage}

\pagestyle{plain}

\settopmatter{printacmref=false}

\author{Damjan Gjurovski}
\email{damjan.gjurovski@cs.rptu.de}
\affiliation{%
  \institution{RPTU Kaiserslautern-Landau}
  \city{Kaiserslautern}
  \country{Germany}
}

\author{Angjela Davitkova}
\email{angjela.davitkova@cs.rptu.de}
\affiliation{%
  \institution{RPTU Kaiserslautern-Landau}
  \city{Kaiserslautern}
  \country{Germany}
}

\author{Sebastian Michel}
\email{sebastian.michel@cs.rptu.de}
\affiliation{%
  \institution{RPTU Kaiserslautern-Landau}
  \city{Kaiserslautern}
  \country{Germany}
}

\begin{abstract}
We propose an advancement in cardinality estimation by augmenting autoregressive models with a traditional grid structure. 
The novel hybrid estimator addresses the limitations of autoregressive models by creating a smaller representation of continuous columns and by incorporating a batch execution for queries with range predicates, as opposed to an iterative sampling approach.
The suggested modification markedly improves the execution time of the model for both training and prediction, reduces memory consumption, and does so with minimal decline in accuracy. 
We further present an algorithm that enables the estimator to calculate cardinality estimates for range join queries efficiently.
To validate the effectiveness of our cardinality estimator, we conduct and present a comprehensive evaluation considering state-of-the-art competitors using three benchmark datasets---demonstrating vast improvements in execution times and resource utilization. 
\end{abstract}

\maketitle

\section{Introduction}
\label{sec:introduction}

Accurate cardinality estimates are a key ingredient to guarantee optimal query execution and to tune applications to reach peak performance. 
Traditional methods based on histograms~\cite{DBLP:conf/sigmod/MuralikrishnaD88, DBLP:journals/vldb/GunopulosKTD05,DBLP:conf/sigmod/GunopulosKTD00, DBLP:conf/sigmod/DeshpandeGR01} or sampling~\cite{DBLP:conf/cidr/LeisRGK017, DBLP:conf/sigmod/0001WYZ16, DBLP:conf/sigmod/ZhaoC0HY18} often face challenges in providing precise estimates, especially for complex queries or skewed data distributions. In response, approaches leveraging machine learning and statistical modeling have gained traction. Supervised algorithms trained on historical data offer the capability to understand complex relationships and patterns~\cite{Fauce, LMKG, MSCN, DuttWNKNC19, LiuXYCZ15, LocalDeepLearningModels, MultiAttributeCE, SimpleModels}. 
Although supervised approaches capture complex attribute correlations for a small memory budget, they typically fail to adapt to query distributions outside of the ones used for training. 
Another innovative field of research comprises models learned directly from the data without requiring query workload generation~\cite{LMKG, Naru, NeuroCard, DeepDB, TupleBubbles, MultiAttributeCE,BayesCard,DBLP:journals/pvldb/HanWWZYTZCQPQZL21}. 
Using probabilistic and deep learning approaches, such as Bayesian networks and autoregressive models, the computation of cardinality estimates for any query workload is possible.

\begin{figure}[!t]
\includegraphics[width=1.\linewidth]{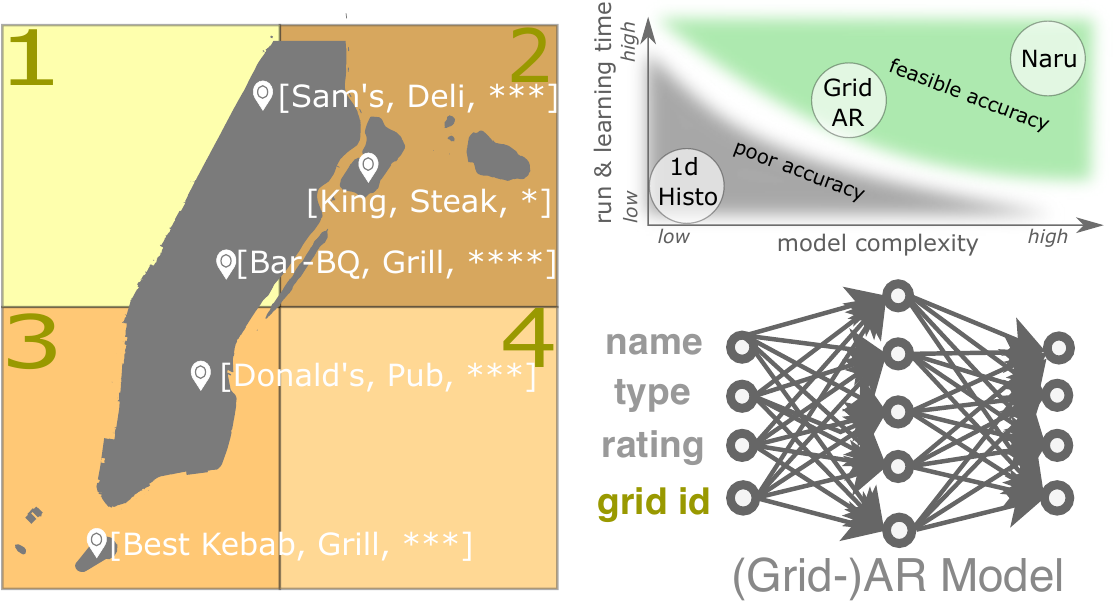}
\caption{
Left and bottom right: 2D four-cell grid over Manhattan's restaurant ratings and AR model with coordinates replaced by grid cells; Top right: space of possible solutions.
}
\label{fig:approachintro}
\end{figure}

When analyzing autoregressive models for cardinality estimation,  we see that the number of distinct values per column in the tables directly impacts the \textit{accuracy}, the \textit{size of the estimator}, and the \textit{prediction speed}. 
The \textbf{accuracy of these models} is directly contingent on the complexity of the data. Consequently, columns that have many distinct values are more likely to possess complex dependencies, making them inherently more challenging to learn. 
Furthermore, deep autoregressive models with many parameters can have substantial \textbf{memory requirements} during training and deployment, limiting their applicability in resource-constrained environments. 
An increase in the dimensionality directly impacts the size of the embedding matrices, i.e., the model grows with the number of distinct values per column. 
Recent efforts to address this problem employ per-column compression~\cite{LMKG, NeuroCard}, which drastically reduces the models' size while it increases the interconnection complexity and leads to a small decrease in accuracy. 
However, even with per-column compression, the overall memory that the estimator occupies can still be improved, as \textit{storing mappings of column values converted to integer values} is required before applying an autoregressive model. The storage of such mappings directly influences the overall memory consumption of the approach.

Another concern with autoregressive models is the poor performance of \textbf{queries with range predicates}. Existing studies~\cite{NeuroCard,Naru} leverage progressive sampling to efficiently sample for answering such queries. However, due to the iterative nature of the estimation process, this approach leads to high execution times. 
Queries involving ranges appear frequently when dealing with temporal, spatial, and metric data. Such queries can be combined with equality predicates on categorical attributes, like in spatial keyword queries. 
\figurename~\ref{fig:approachintro} illustrates such a scenario, highlighting points of interest like restaurants. Predicates on categorical attributes, such as restaurant names or types, typically check for equality, while  attributes like restaurant location and opening hours are often used with range predicates. 

\noindent{\textbf{To overcome the problems of autoregressive models, we propose Grid-AR~\footnote{Code available at: 
\url{https://github.com/dbislab/Grid-AR}
}}}, 
a hybrid structure that integrates a \textit{grid structure} for columns typically used in range predicates and an \textit{autoregressive model} for columns used in equality predicates.
The grid acts as a preliminary filter for predicates, leading to the segmentation of data into simpler partitions that substitute the continuous attributes in the original data before the autoregressive model is trained. 

Doing so promises several \textbf{benefits}:
First and foremost, it allows fast prefiltering of tuples that need to be considered by pruning the search space upfront.
Secondly, by substituting the ample space of possibilities with prefiltered partitions, the need for progressive sampling during estimation is completely circumvented, drastically speeding up the execution.
Finally, such a hybrid structure avoids storing large embeddings and many dictionary mappings for the numerical columns, resulting in a smaller memory consumption.

Mixed data containing categorical and continuous attributes is often queried with \textbf{range join queries}. 
An example is locating restaurants of type \texttt{deli} with better ratings than the restaurants of type \texttt{pub}. Such a query will result in a self join over restaurants with a join condition over the rating column. 
Range joins between two relations typically narrow down to three types of conditions: inequality, a point in interval, and interval overlap conditions. 
Related work~\cite{Hyper, DuckDB, DBLP:journals/pvldb/ReifN22} optimizes such queries for a complete query execution rather than cardinality estimation.
\textbf{Therefore, we introduce a generalized definition of range join queries and propose an algorithm using the power of Grid-AR to provide fast and accurate cardinality estimates for range join queries.}

\subsection{Problem Formulation}
\noindent{\underline{\textbf{Single table estimates:}}}
Let $R$ be a relational table with $n$ columns $c_1, c_2, ..., c_n$ and $Q(R)$ a query involving conjunction of predicates over any column of $R$, i.e., $Q(R) = \bigwedge\limits_{i} {R.c_i\:  \theta\:  v}$, where $\theta\in\{=, >, <, \ge, \le\}$ and $v$ is a value usually in the domain $D(c_i)$ of $c_i$. Let $|Q(R)|$ denote the cardinality, i.e., number of tuples qualifying for the query. The objective is to attain an accurate estimate of $|Q(R)|$, considering metrics such as the $q\!-\!error$.

\noindent{\underline{\textbf{Range join estimates:}}} 
Let $R_1, R_2, ..., R_n$ be relational tables with an arbitrary number of columns. Let $Q$ be a query that considers a join between the two tables of the form $Q(R_1, R_2) = \bigwedge\limits_{i, j} {R_1'.c_i\: \theta\: R_2'.c_j}$, where $\theta\in\{>, <, \ge, \le\}$, and $R_1'$ and $R_2'$ are prefiltered tables with already applied equality and range predicates.
The aim is to achieve an accurate estimation of multi-table joins $|Q(R_1, R_2)... Q(R_{n-1}, R_n)|$.

\subsection{Contributions and Outline}
The main contributions of this paper are as follows:
\begin{enumerate}
	\item We propose a cardinality estimator, termed \textbf{Grid-AR}, by combining a grid and an autoregressive structure (Section~\ref{sec:grid-ar-approach}). 
	\item We develop an efficient algorithm for fast execution of queries over single tables through employing grid structure prefiltering while reducing the memory overhead of storing additional per-column value mappings (Section~\ref{sec:grid_ar_single_table_query}).  
	\item We present an algorithm employing Grid-AR for estimating the cardinality of range join queries (Section~\ref{sec:grid_ar_range_join_query}).
	\item We perform a comprehensive experimental study over one synthetic and two real-world datasets (Section~\ref{sec:experiments}).
\end{enumerate}

To the best of our knowledge, this is the \textit{first approach combining learned and non-learned approaches for estimating cardinalities of range joins}.  
Thus, with Grid-AR, we extend the possibilities of using 
learned models for a task that has only scarcely been addressed for relational databases so far.


\section{Preliminaries}

\subsection{Grid Index}

A grid index is a data structure used in databases and Geographic Information Systems to organize and query multi-dimensional data~\cite{SametBookMultiDimIndex}. 
It provides a coarse-grained partitioning of space, useful for quickly eliminating irrelevant data portions upon point, nearest neighbor, or range queries. 
Each cell in the grid contains pointers to multi-dimensional points that intersect or are contained within that cell.
While often used for two-dimensional points, the grid index is also efficient for more dimensions.

\noindent{\underline{\textbf{Analysis \& Goals:}}}
Unlike the original intention of utilizing the structure for indexing and querying actual points, our objective is to adapt the structure for estimation. 
Although our algorithms are compatible with 
any alternative learned or non-learned indexing structure that can handle more than one dimension and group the data into regions~\cite{SametBookMultiDimIndex,Flood,DBLP:conf/vldb/PoosalaI97}
,
we opt for the grid index due to its fast processing speed and low memory consumption.
While a tree-like structure may provide a better grouping of the points, a uniformly created grid can return the cell containing a query point in constant time.  
However, the efficiency of a grid index can be affected by issues such as data skew and varying object densities across the cells. Thus, we explore dynamically adapting grid cell boundaries rather than adhering to a uniform division.

\subsection{Autoregressive Model}

An autoregressive model is an unsupervised model that can estimate a joint distribution $P(x)$ from a set of samples.
For a predefined attribute ordering, the autoregressive property is that the output of the model contains the density for each attribute conditioned on the values of all preceding attributes. 
Given as input $x = [x_1, x_2, \ldots x_d]$, the autoregressive model produces $d$ conditional probabilities. The product of the conditionals results in the point density $P(x)$.

\noindent Formally, for $x = [x_1, x_2, \ldots, x_d]$, the probability is calculated as:
\begin{equation}
P(x)\!=\!P(x_1)P(x_2|x_1)\ldots P(x_d|x_1, \ldots , x_{d-1})\!=\!\prod_{d=1}^{D} P(x_d|x_{<d})
\end{equation}
In the context of relational databases, Naru~\cite{Naru} considers that each input $x$ corresponds to one tuple of a table. The approach uses point density estimates $P(x)$ and conditional densities $P(x_i|x_{<i})$ obtained through various network architectures, including masked multi-layer perceptrons~\cite{MADE,ResMADE} and masked self-attention networks~\cite{Transformer}. The autoregressive models capture complex dependencies in relational data by utilizing information masking during the encoding process. Encoding and decoding strategies are tailored to the size of the domain, with one-hot encoding for small and embeddings for larger domains. Further improvement~\cite{NeuroCard} employs per-element compression with embedding for further reducing the space. The querying process in Naru is facilitated by the introduction of \textit{Progressive Sampling}, an unbiased estimator for range predicates that focuses on high-density regions by employing an iterative algorithm. 
Additionally, an optimization known as wildcard-skipping allows for fast execution of queries over a subset of the columns.

\noindent{\underline{\textbf{Analysis \& Goals:}}}
 While many alternative data-driven models, such as Bayesian networks exist,  the results of \citet{Naru} suggest that autoregressive models provide better accuracy.
Although highly accurate, autoregressive models come with two main flaws. 
Queries with range predicates over autoregressive models are based on an iterative process, i.e., progressive sampling. Sampling yields large execution times and is negatively influenced by the increase in the number of columns used in query predicates. 
For instance, upon analyzing the execution over the \textit{Flight} dataset, we have observed that \textbf{doubling the number of query predicates} from $2$ to $4$ results in almost \textbf{doubling the average estimation time}, i.e., $440$ to $970$, and $390$ to $860$ milliseconds for the models without and with compression, respectively.  
Moreover, due to the nature of machine learning models, the input to the model must be an integer. Hence, 
we also necessitate an integer mapping of the values. 
Although appropriate for categorical data, employing a dictionary mapping leads to substantial memory consumption in instances involving continuous data with high skewness.
For the \textit{Flight} dataset, as expected, the total size of all dictionaries increases linearly with the number of columns used in the estimator.
We aim to fine-tune the estimator's structure to tackle an inherent challenge associated with progressive sampling in queries with range predicates, all while maintaining a compact memory footprint.

\section{Grid-AR}
\label{sec:grid-ar-approach}

\begin{figure*}[!t]
\centerline{\includegraphics[width=1.\linewidth]{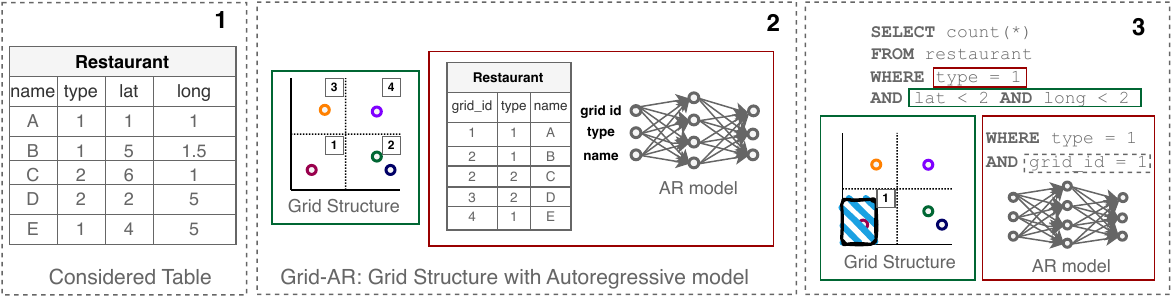}}
\caption{Creation and querying of Grid-AR. Creation of the grid structure for the continuous attributes and an autoregressive model for the categorical ones together with the assigned grid cell id (part 2). For an example query, the Grid-AR parts responsible for the different types of query predicates are used for producing cardinality estimates (part 3).}
\label{fig:grid-ar} 
\end{figure*}

To ease both the creation and the querying over ranges of the autoregressive model for the task of cardinality estimation, 
we propose a novel hybrid structure called \textbf{Grid-AR}.  
Consider a table $T$ with columns $C = \{c_1, c_2, \ldots, c_n\}$. We distinguish between two \underline{disjoint} sets of attributes:
\begin{itemize}
\item $CR = \{cr_1, cr_2, \ldots, cr_k\}$ where $CR \subseteq C$ are columns typically involved in range predicates.
\item $CE = \{ce_1, ce_2, \ldots, ce_l\}$ where $CE \subseteq C$ are columns typically involved in equality predicates.
\end{itemize}
This way of organizing the columns is based on an observation that columns that store continuous data, e.g., temporal or spatial data, are likely to be associated with range predicates, i.e., filtering by a range of values. Conversely, text attributes are more likely to be included in equality predicates, implying exact matches. First, we build a grid based on the $CR$ columns, and then, using the grid cells together with the $CE$ columns, we train the autoregressive model. 

\subsection{The Grid Structure}
\label{ssec:grid_index_creation}

To create the \textbf{|CR|-dimensional} grid structure of Grid-AR, we utilize the values of the columns in $CR$ to superimpose a grid on the data. 
For each dimension (column), we split the data into $m$ buckets. To adequately capture the diverse characteristics of individual columns, the number of buckets differs for each column. A grid cell is represented by the boundaries of the buckets of every dimension. We consider two ways of creating buckets, one by simply following the uniformity assumption and the other by considering a given per-column data distribution. In the following, we will use the terms columns and dimensions interchangeably to refer to the columns of the table that represent the dimensions of the grid structure.

\noindent{\underline{\textbf{Uniform grid:}}} In the first option, we group the column values, assuming the uniformity assumption. More specifically, within each dimension, the grid is partitioned into evenly spaced buckets spanning from the minimum to the maximum value. For every dimension $i$, there will be $m_{cr_i}$ buckets that the column values will be organized into. The size of each bucket  
is computed as $bucket\_size_{cr_i} = \frac{max(vals_{cr_i}) - min(vals_{cr_i}) + 1}{m_{cr_i}}$, where $vals_{cr_i}$ are the distinct values of the column $cr_i$. 
Thus, a column $cr_i$ with value $val_{cr_i}$ will be mapped to a bucket based on the equation 
$bucket_{cr_i} = \left \lfloor\frac{val_{cr_i} - min(vals_{cr_i})}{bucket\_size_{cr_i}}\right \rfloor$. 
A tuple from the table is matched to a grid cell by considering the buckets to which the values of the tuple belong. This way of creating the grid structure requires storing only a few necessary information: the min and max value per dimension, the buckets' boundaries per dimension, the buckets' size, and for every grid cell, the min and max value per dimension for the qualifying tuples. 

\noindent{\underline{\textbf{CDF-based grid:}}} As an alternative to the grid creation above, we consider a grid tailored to the data distribution based on the cumulative distribution function (CDF) for every dimension. 
Instead of dealing with a direct value for each column $val_{cr_i}$, we consider the model output $f(val_{cr_i})$ representing the CDF of the value $val_{cr_i}$. In other words, the model responsible for computing the per dimension CDF will receive as input the value $val_{cr_i}$ and will output the fraction of points with values that are less or equal to $val_{cr_i}$. 
In Grid-AR, we utilize a decision tree regressor~\cite{DBLP:books/wa/BreimanFOS84} as a powerful tool for statistical modeling and learning the distribution of every column of the table. Unlike traditional decision trees for classification, regression trees are designed to predict continuous numerical outcomes. The construction of a regression tree involves recursive splitting of the dataset based on selected features and corresponding thresholds, ultimately leading to decision nodes and leaf nodes. At each decision node, the dataset is partitioned into subsets, guided by criteria such as mean squared error or variance reduction, with the goal of minimizing the variance of the target variable within each subset. The model accuracy improves by incrementing the number of levels; however, this increases the estimation time. 
The variance for a node \(n\) is calculated as \(\text{Var}(D_n) = \frac{1}{|D_n|} \sum_{i \in D_n} (y_i - \bar{y}_n)^2\), where \(y_i\) is the target variable, \(|D_n|\) is the data count, and \(\bar{y}_n\) is the mean of \(y\) in node \(n\). Predictions at leaf nodes (\(\hat{y}_n\)) are determined as \(\hat{y}_n = \frac{1}{|D_n|} \sum_{i \in D_n} y_i\). Model accuracy is assessed through the mean squared error over the actual and predicted value. 
These equations capture the essence of regression trees, highlighting their role in effective data partitioning, prediction, and model evaluation. For the grid structure of Grid-AR, we need to store the information that makes up the regression trees, the number of buckets per dimension, together with the minimal and maximal value for every dimension in all grid cells. For Grid-AR, a column $cr_i$ with a value $val_{cr_i}$ will be mapped to a bucket based on the equation: 
$$bucket_{cr_i} = \left \lfloor f_{cr_i}(val_{cr_i})\right\rfloor * m_{cr_i}$$ 
where $m_{cr_i}$ is the number of buckets that the values of $cr_i$ will be grouped into and $f_{cr_i}$ approximates the $CDF$ value of $val_{cr_i}$.

No matter the choice of the grid structure, a tuple $t$, considering only the columns in $CR$, corresponds to a grid cell id identified with 
$gc_{id} = \left[bucket_{cr_1}, \ \ldots, \ bucket_{cr_k}\right]$.  
The organization of cells and the generation of grid cell ids is carried out through a depth-first traversal of cells along the dimensions.

\textbf{CDF grids, in comparison to uniform grids,} create splits according to the data distribution. The resulting split boundaries are not equally spaced across the minimum and maximum values of the dimension but are instead based on the cumulative distribution function (CDF). 
Consequently, regardless of the data distribution, a CDF grid ensures that each bucket contains roughly the same number of points, with more splits in dense regions and fewer in sparse regions. In contrast, uniform grids might be simpler to implement and computationally less complex, but their performance degrades with highly skewed or non-uniform data distributions, which results in buckets with a disproportionately large number of points, leading to increased errors.

\noindent{\underline{\textbf{Number of grid cells:}}} In Grid-AR we can independently set the number of buckets $m_{cr_i}$ for every dimension. Nevertheless, choosing a feasible number of buckets depends on the considered column and its values, while an optimal choice is clearly dependent on accurate assumptions on the query workload. 

Determining the \textbf{number of buckets per dimension} means balancing a trade-off: When the grid has fewer buckets, the size of the estimator will be smaller, and the estimate can be produced faster since less cells are considered for answering the query. However, this comes at the cost of a lower accuracy as cell boundaries per dimension will be coarser, including tuples that potentially do not match the query predicate. In contrast, when the number of buckets is increased, the ranges are more tightly bound. Hence, for the same predicate more cells are considered, which increases the estimation time but captures the query predicates better.

In our experimental evaluation (Section~\ref{ssec:exp_model_analysis}), we show the effects of varying the number of grid cells. To achieve an optimal balance between the execution time, memory, and accuracy, for smaller datasets where a clear pattern in the grouping of the column values can be observed, we rely on the analysis of per-column values. However, when we cannot draw any conclusions about the number of buckets per column directly from the dataset, we apply a cost model for identifying the grid layout, similar to the one proposed by Nathan et al.~\cite{Flood}. We consider a sample query workload, and based on it, try to identify the optimal number of ranges per dimension to achieve the right balance between the time, memory, and accuracy.

\subsection{The Autoregressive Model}
\label{ssec:ar_model_creation}
In the \textbf{second step} of the creation of Grid-AR, we proceed with creating the autoregressive model. To perform the training, the autoregressive model, in addition to the $CE$ columns, requires as input the grid cell ids. In other words, a new tuple is created by combining the grid cell id to which the tuple belongs together with the $CE$ attributes of the tuple. The table is thus transformed from $cr_1, \ldots, cr_k, ce_1, \ldots, ce_l$, to $gc_{id}, ce_1, \ldots, ce_l$.  
With a designated column order, the autoregressive model is trained to comprehend the interdependent relationship between the values of a column and the values of the preceding columns in the specified order. Consequently, in this scenario, the model acquires an understanding of the dependencies between the grid cell ids and the remaining columns. As already outlined, to make the autoregressive model applicable to highly heterogeneous columns, we apply a per-column lossless compression~\cite{LMKG}. A column qualifies for compression if it has more than a prespecified number of distinct values $\gamma$. The hybrid structure brings forth an important advantage. Using the grid structure, we avoid the storage of mappings of continuous columns to consecutive integers. This directly speeds up the model and drastically reduces the memory footprint. 

\noindent{\underline{\textbf{Example:}}}
In \figurename~\ref{fig:grid-ar}, we demonstrate the creation of our hybrid structure Grid-AR when considering information about restaurants depicted in the \texttt{Restaurant} table. We first distinguish between attributes that can be used in range predicates, i.e., continuous attributes, and attributes that can be used for equality predicates, i.e., textual attributes. Thus, for constructing the grid, we establish a grid pattern over the location coordinates \texttt{lat} and \texttt{long}. For the sake of clarity, in this example, we partition each dimension into two regions, yielding a total of four grid cells. We assign a $gc_{id}$ to every cell which is created from the per-dimension bucket ids. The grid cell id, together with the textual attributes, is later used for training the autoregressive model.
For instance, restaurant \textit{A} has the coordinates $(1, 1)$ which align with the grid cell with id $1$. As depicted in part $2$ of \figurename~\ref{fig:grid-ar}, the grid cell ids will be assigned to the respective tuples based on the columns used for creating the grid structure; in this example, the columns \texttt{lat} and \texttt{long}.

\section{Grid-AR Single Table Queries}
\label{sec:grid_ar_single_table_query}

\begin{algorithm}[!t]
\centering
\caption{Grid-AR single table query}
\label{alg:single_table_query}
\begin{algorithmic}[1]

\Function{estimate}{\textit{Query} $Q$, \textit{Grid} $GRID$, \textit{AR Model} $AR$}
\State{$Q_{grid}, Q_{AR} \gets split(Q)$} \Comment{split $Q$ according to columns}
\State{$GC_{grid} \gets cells\_for\_query(Q_{grid}, GRID)$}\Comment{query grid}
\State{$est_{AR} \gets AR.estimate(GC_{grid}, Q_{AR})$}\Comment{query AR-model}
\State{$sel \gets 0$}
\For{$gc_i$ $\in$ $GC_{grid}$}\Comment{for each qualifying grid cell}
\State{$est_{AR_i} \gets est_{AR}[gc_i.id]$}
\State{$est_{gc_i} \gets cell\_estimate(gc_i, Q_{grid})$}
\State{$sel \gets sel + est_{gc_i} * est_{AR_i}$}
\EndFor
\State{\textbf{return} $sel$}
\EndFunction

\Function{cells\_for\_query}{\textit{Query} $Q$, \textit{Grid} $GRID$}
\State{$GC_{grid} \gets \{\}$}
\For{$dim$ $\in$ $Q.dims$}\Comment{for each dimension}
\If{$dim == 0$}
\State{$GC_{grid} \gets GRID.gcs\_per\_dim(Q[dim])$}
\Else
\State{$GC_{grid} \gets GC_{grid} \cap GRID.gcs\_per\_dim(Q[dim])$}
\EndIf
\EndFor
\State{\textbf{return } $GC_{grid}$}
\EndFunction

\Function{cell\_estimate}{\textit{Grid cell} $gc$, \textit{Query} $Q$}
\State{$b_{min} \gets max(gc.{min}, Q.{min})$}\Comment{min for each dimension}
\State{$b_{max} \gets min(gc.{max}, Q.{max})$}\Comment{max for each dimension}
\State{$volume \gets 1$}
\For{$dim$ $\in$ $b_{min}.dims$}\Comment{for each dimension}
\State{$volume \gets volume * (b_{max}[dim] - b_{min}[dim])$}
\EndFor
\State{\textbf{return} $volume/gc.volume$}
\EndFunction

\end{algorithmic}
\end{algorithm}

During querying, Grid-AR estimates the query selectivity using the estimates for the $CR$ columns on the grid structure and the estimates for the $CE$ columns on the autoregressive model. Algorithm~\ref{alg:single_table_query} illustrates the querying process of Grid-AR for providing the cardinality for queries over single tables involving both range and equality predicates. When querying the structure, the algorithm divides the query into two subqueries: one involving columns used in the grid structure and the other for the autoregressive model, i.e., $Q_{grid}$ and $Q_{AR}$ (Algorithm~\ref{alg:single_table_query}, Line $2$). 

The first phase of the algorithm involves using the grid to identify cells overlapping with the query boundaries in $Q_{grid}$ (Algorithm~\ref{alg:single_table_query}, Line $3$). This is done by obtaining the grid cells that intersect with the query boundaries in each dimension (Algorithm~\ref{alg:single_table_query}, Lines $11$--$18$). Let $GC_i$ represent the set of cells  that overlap with the query boundaries in the $i$-th dimension obtained with the method $gcs\_per\_dim$. The intersection is then applied to these sets to retain only the grid cells that qualify in all dimensions:
$GC_{grid} = GC_1 \cap GC_2 \cap \ldots \cap GC_k$.
In the end we obtain $GC_{grid}$ representing the filtered set of grid cells that satisfy $Q_{grid}$.

Once the qualifying cells have been determined, 
the algorithm switches to the autoregressive model. In this step, the model receives the ids of the cells in $GC_{grid}$ along with the remaining values for the columns in $CE$ for which the model is responsible (Algorithm~\ref{alg:single_table_query}, Line $4$).
More specifically, for each grid cell, the estimate is computed as:
$$est_{AR} = P(gc_{id}, ce_1, ..., ce_l) = \\ P(gc_{id}) ... P(ce_l|ce_{i<l}, gc_{id})$$ 
\textbf{Note that} the estimation is a simple point density estimation in the autoregressive model, computed in a single forward pass in a batch for all the grid cells. This way of computation circumvents the need for multiple iterations when sampling with the progressive sampling estimation algorithm. As a result, we get an estimate for the cardinality of the query $Q_{AR}$ for each grid cell.

The algorithm proceeds to estimate the query selectivity by considering the results from each cell (Algorithm~\ref{alg:single_table_query}, Lines $6$--$9$).
The calculation takes into account the cardinality of the autoregressive model and the grid cell selectivity.
In the grid structure, the algorithm estimates the \textit{overlap volume} between the query $Q_{grid}$ and the grid cell boundaries for every dimension, detailed in the function called \texttt{cell\_estimate} (Algorithm~\ref{alg:single_table_query}, Lines $19$--$25$).
We return the fraction of overlap by dividing it with the volume of the whole grid cell, i.e., $est_{gc_i} = \frac{V(gc_i \ \cap \ Q_{grid})}{V(gc_i)}$.
As the query boundaries do not always completely contain a grid cell, \textbf{the overlap} indicates the proportion of grid cell points that should be included according to the query boundaries. Hence, if the query boundaries overlap with only $50\%$ of the grid cell volume, the overlap value will be $0.5$.

The volume obtained from the grid cell and the cardinality of the autoregressive model are multiplied and returned as the final estimate for the grid cell $est_{gc_i} * est_{AR}$ (Algorithm~\ref{alg:single_table_query}, Line $9$). The individual estimates are summed up and returned as a result. 
The worst-case complexity of the single table estimate is
$O(|GC_{grid}| * Cost_{AR})$ where $Cost_{AR}$ is a forward pass cost and depends on the number of layers of the autoregressive model.

\noindent{\underline{\textbf{Example:}}}
\figurename~\ref{fig:grid-ar} shows the execution of a simple query over the table \texttt{Restaurant} that filters the coordinates as well as the type of the restaurant. In the first step, we use the grid structure to filter out the possible grid cells that qualify for the range predicate. In this example, this will be the grid cell $1$. The grid cell, together with the equality predicate $type = 1$, is forwarded to the model. The final estimate is calculated by using the overlap percentage of the query and the qualifying cells as well as the estimate $P(gc_{id} = 1, type = 1) = P(gc_{id} = 1)*P(type = 1|gc_{id} = 1)$.

Querying Grid-AR brings one significant benefit over a textbook autoregressive model. The computation of range predicates in the autoregressive model through progressive sampling requires $n$ iterations of forward passes for each of the $n$ columns in the query. With Grid-AR, the iterative computation is no longer necessary as the approach no longer uses progressive sampling. 
We will further demonstrate the effects on the execution time in our experiments.

\section{Grid-AR Range Join Queries}
\label{sec:grid_ar_range_join_query}

The creation of the hybrid Grid-AR structure facilitates the execution of another type of query commonly performed on data containing continuous attributes, namely \textbf{range join queries}. Existing research~\cite{DuckDB, Hyper, DBLP:journals/pvldb/ReifN22} concentrates on enhancing the execution of such joins and overlooks the aspect of estimating their cardinality. To the best of our knowledge, we are the \textit{first to perform estimation over such joins utilizing the potential of learned models}.

A range join is an operation where two relations are joined based on an inequality, a point in interval, or interval overlap condition. 
An \textbf{inequality join} is the basic query form, where two attributes from two relations are compared to one another with any comparison operator except for equality. 
A \textbf{point in interval range join} contains a join condition characterized by predicates specifying that a value from one relation falls within the interval defined by two values from another relation. 
Let $R$ and $S$ be two relations. The point in interval join condition is expressed as $R.value \in [S.lower, S.upper]$, where $R.value$ is an attribute in $R$, and $S.lower$ and $S.upper$ define the lower and upper bounds of the interval and are values from an attribute in $S$.  
An \textbf{interval overlap range join} is a join operation in relational databases, defined by conditions expressed through predicates that require an overlap of intervals between two values, each derived from two relations. For two relations $R$ and $S$ this join can be expressed as:
$ R.lower \leq S.upper \, \text{AND} \, R.upper \geq S.lower $ where $ R.lower $ and $R.upper$ are the lower and upper bounds for an attribute in the first relation, and $S.lower$ and $S.upper$ are the corresponding bounds in the second relation. 

We generalize these queries into a more generic form. In one query, we allow multiple join predicates of the form:  
$f(R.c_i)\: \theta\: g(S.c_j)$ such that $\theta \in \{\leq, <, \geq, >\}$ and $f$ and $g$ are arbitrary expressions performed over the attribute values. The expression $f(R.c_i)$ can range from a simple injective function $f(R.c_i) = R.c_i$ to a more elaborate one such as $f(R.c_i) = R.c_i * 2 + 100$.

\begin{algorithm}[t]
\centering
\caption{Grid-AR range join query}
\label{alg:range_join_query}
\begin{algorithmic}[1]
\Function{range\_join\_est}{\textit{Query} $Q$, \textit{Table} $R$, \textit{Table} $S$}
\State{$Q_{l}, Q_{r}, Q_{j} \gets split(Q)$}\Comment{split $Q$ according to columns}
\State{$GC_{l} \gets Q_{l}(R)$}\Comment{query estimator of $R$}
\State{$GC_{r} \gets Q_{r}(S)$}\Comment{query estimator of $S$}
\State{$card \gets 0$}
\State{$GC_{r\_card} \gets cells\_card\_array(GC_r)$}\Comment{per-cell cardinality}

\For{$gc_{li}$ $\in$ $GC_{l}$}\Comment{for each grid cell qualified for $Q_{l}$} 
\State{$op \gets init\_arr(gc_{li}.card, size=|GC_r|)$}

\For{$j_c$ $\in$ $Q_{j}$}\Comment{for each join condition}
\State{$GC_{r} \gets get\_sorted(GC_{r}, j_c)$}\Comment{sort according to $j_c$}
\For{$gc_{rj}$ $\in$ $GC_{r}$} \Comment{for each cell qualified for $Q_{r}$}
\State{$cond_{fulfill} \gets check\_condition(j_c, gc_{li}, gc_{rj})$}
\If{$cond_{fulfill} == 0$}\Comment{unsatisfied}
\State{$op[gc_{rj}.id] \gets 0$}
\ElsIf{$cond_{fulfill} == 1$}\Comment{satisfied}
\State{\textbf{break}}
\Else \Comment{partially satisfied}
\State{$op[gc_{rj}.id] \gets op[gc_{rj}.id] * p(j_c, gc_{li}, gc_{rj})$}
\EndIf
\EndFor
\EndFor
\State{$card \gets card + sum(op * GC_{r\_card})$}
\EndFor
\State{\textbf{return } $card$}
\EndFunction

\end{algorithmic}
\end{algorithm}

\subsection{Estimation Algorithm}
Grid-AR is, by design, set up to perform cardinality estimation for range joins, as all the continuous attributes are part of the grid structure.
We split the cardinality estimation in \textbf{two steps}. 

\noindent{\underline{\textbf{Grid-AR query phase:}}} First, we split the query into subqueries, isolating the predicates over individual tables from those related to the join conditions (Algorithm~\ref{alg:range_join_query}, Line $2$). These segmented subqueries $Q_l$ and $Q_r$ are then processed over the individual estimators for the respective table. Upon executing a query with a single table estimator, instead of providing a final cardinality for the entire table, we return the cardinality of the query for each grid cell (Algorithm~\ref{alg:range_join_query}, Lines $3$--$4$). 
Since the per-table subqueries already filter out some of the grid cells that do not qualify for the subquery result,  
there will be a significantly smaller number of cells than the total number of grid cells in the estimator. Consequently, the initial prefiltering is carried out.
The algorithm that performs computation over the individual tables is as described in Algorithm~\ref{alg:single_table_query}.

\noindent{\underline{\textbf{Range join query phase:}}} The second part uses the estimates produced by Grid-AR and the grid cells produced from the individual tables to estimate the cardinality for the range join query (Algorithm~\ref{alg:range_join_query}, Lines $6$--$19$). 
Let $gc_{l1}, gc_{l2}, \ldots$ be the result grid cells from the left table $R$ and $gc_{r1}, gc_{r2}, \ldots$ be the result grid cells from the right table $S$.
We iterate through each $gc_{li}$, and for each $gc_{rj}$, we ascertain if and to what extent the conditions specified in the range join are satisfied. 
\textbf{Note that} the grid cells $GC_{r}$ of $S$ are sorted according to the join condition to allow an early termination (Algorithm~\ref{alg:range_join_query}, Line $10$). Furthermore, since the calculations for a pair of cells are independent, we parallelize the algorithm with respect to the grid cells from table $R$.
For each condition, if we have an expression over the values involved in the join, we apply the expression to the boundaries of the buckets for the dimension on which the expression appears. We then distinguish between \textit{three cases}. 

\numcircledmod{1} If the grid cell $gc_{rj}$ does not satisfy at least one join predicate condition, we do not consider the grid cell estimate in the final cardinality (Algorithm~\ref{alg:range_join_query}, Line $13$).
\numcircledmod{2} Conversely, if all of the boundaries of the grid cell $gc_{rj}$ completely satisfy all join conditions when compared to $gc_{li}$, we consider a complete predicate fulfillment (Algorithm~\ref{alg:range_join_query}, Line $15$). 
In the best-case scenario, when the per-table estimates are exactly correct, then the previous two computations will result in the actual correct cardinality. 
\numcircledmod{3} The grid cell $gc_{rj}$ partially satisfies the conditions specified in the range join when compared to the cell $gc_{li}$ (Algorithm~\ref{alg:range_join_query}, Line $18$).
In such case between two grid cells, for each condition $\theta$, we calculate the probability that the $r$-th condition is satisfied as $op_r$. 
The same calculation is performed for all of the conditions in the join query in the end, resulting in $k$ individual condition satisfaction probabilities $op_1, \ldots, op_k$. All of the probabilities are then multiplied, resulting in the overall predicate conditions probability of the whole range join query for the considered two grid cells. Intuitively, when the condition is completely not satisfied, then $op_r = 0$, and when the condition is completely satisfied, then $op_r = 1$. 

The final estimate for the cardinality of two cells $gc_{li}$ and $gc_{rj}$ is $card_i * card_j * op_1 \ldots * op_k$, where $card_i$ and $card_j$ are the estimated cardinalities of $gc_{li}$ and $gc_{rj}$, respectively. 
To produce the range join estimate, we iterate through all cells and sum over the estimated cardinality (Algorithm~\ref{alg:range_join_query}, Line $19$), calculated as: 

$$card = \sum_{i=1}^{n} \sum_{j=1}^{m} card_{i} * card_{j} * \prod_{r=1}^{k} op_{ijr}$$
where $n$ and $m$ are the number of grid cells produced as a result of the subquery of the left table and right table, and $k$ is the number of conditions in the range join.

\noindent{\underline{\textbf{Join predicate fulfillment:}}} 
For the join conditions between $R$ and $S$, we define the probability of a condition $\theta$ being satisfied for two grid cells as $op = p(x<y)$ or $op = p(x>y)$, i.e., the probability that a variable $x$ is smaller (or larger) than $y$, where $x$  
$\in [x_{min}, x_{max}]$ and $y$  
$\in [y_{min}, y_{max}]$. The ranges correspond to the cell boundaries for the columns considered in the join condition. 

To calculate $op$, we explore several variants. 
One way of computation is through \textit{sampling}. Given $n$ samples $S_x = \{s_{x1}, \ldots, s_{xn}\}$ for $x$ and $n$ samples $S_y = \{s_{y1}, \ldots, s_{yn}\}$ for $y$, where $\forall{s_{xi}}: s_{xi} \in [x_{min}, x_{max}]$ and $\forall{s_{yi}}: s_{yi} \in [y_{min}, y_{max}]$, we calculate $op$ as: $$op = \frac{|S_x \:  \theta \:  S_y|}{n}$$   
which represents the fraction of samples that satisfy the condition divided by the number of retrieved samples $n$.
Alternatively, we can use \textit{double integration} to determine the probability $op$ of a condition being met. We observed that both approaches yield comparable probabilities when increasing the number of samples. Due to the computational efficiency, we chose to proceed with sampling.

The worst-case complexity of a two-table join is $O(|GC_l| * |GC_r| * |Q_j|)$, which occurs only when the join conditions necessitate that every grid cell from one table matches with every grid cell from the other table, otherwise, the algorithm skips most grid cells.

\begin{figure}[!t]
\centerline{\includegraphics[width=.95\linewidth]{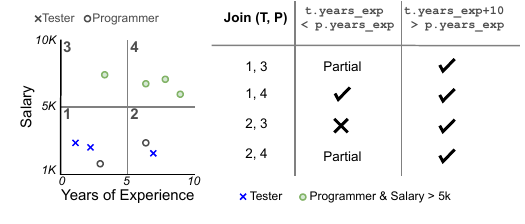}}
\caption{
Example range join query. The grid structure with values is shown on the left-hand side and the comparison between qualifying grid cells on the right-hand side. 
}
\label{fig:range_join_query_example} 
\end{figure}

\noindent{\underline{\textbf{Range join query example:}}} To clarify the execution of our algorithm, we consider the example depicted in \figurename~\ref{fig:range_join_query_example} featuring data points from the \texttt{Employee} table.
The grid structure for the continuous attributes of the \texttt{Employee} table is depicted on the left-hand side of \figurename~\ref{fig:range_join_query_example}. To illustrate the estimation process, we consider a self-join of the following form: 
\begin{Verbatim}[tabsize=2]
	SELECT COUNT(*)
	FROM Employee t, Employee p
	WHERE t.job = "Tester" AND p.job = "Programmer" 
	AND p.salary > 5000 AND t.years_exp < p.years_exp 
	AND t.years_exp + 10 > p.years_exp 
\end{Verbatim}
where we join employees that are \textit{Testers} and employees that are \textit{Programmers} with salary $> 5000$ based on years of experience.

A query that involves a join between the two tables requires two Grid-AR estimators, but for this example, they will have the same structure, as depicted in \figurename~\ref{fig:range_join_query_example}. Initially, we evaluate the two distinct subqueries over the respective table estimator to retrieve the qualifying grid cells. Thus, for the predicate \texttt{t.job = "Tester"}, the first and the second grid cells qualify.
For the predicates \texttt{p.job = "Programmer"} and \texttt{p.salary > 5000}, the third and the fourth grid cells qualify. Alongside the qualifying cell, we also obtain the estimate of the cardinality of each cell given the  
predicates.
Subsequently, we iterate through the cell combinations and assess whether each condition is satisfied.
To optimize the execution, we sort the grid cells from the second table according to the dimension in the current considered condition. Thus, in this example, we will sort the cells in the order $3, 4$. For the first join condition, when comparing the grid cell $1$ of the \textit{Testers} and the grid cell $3$ of the \textit{Programmers}, we need to compute the probability of the condition being satisfied due to the overlap in their ranges. However, when comparing grid cell $1$ with $4$, we can ascertain that every point from cell $1$ is undoubtedly smaller than any point in cell $4$ since the entire range of cell $1$ is smaller than that of cell $4$. Thus, the probability that the condition is satisfied is $op = 1$. In contrast, the entire cell $2$ is completely larger than cell $3$, leading to $op=0$. 
We follow a similar process for the other condition, where the optimization from sorting becomes salient. 
For the considered condition, we can exploit the sorting order to skip the comparison between some cells. 
In this case, the cell order is $4,3$. Recognizing that cell $1$ is already larger than cell $4$ when assessing the conditions, it implies that all subsequent grid cells in the order will also be smaller. Thus, $op$ is set to $1$ without having to compare further.

\noindent{\underline{\textbf{Multi-Table Join Estimation:}}}
While multi-table range joins are not typical, our proposed algorithm supports them.
More specifically, a \textit{chain multi-table} join is broken down into several range joins involving pairs of relations., i.e., $(Q_1, R_1, R_2), \ldots, (Q_{n-1}, R_{n-1}, R_n)$. The processing of the complex join query starts from the first join pair by executing Algorithm~\ref{alg:range_join_query}. However, as an output, instead of the total cardinality, the estimated cardinality is accumulated for each cell $gc_j$ of $R_2$ by summing up all join cardinalities with the corresponding grid cell partners from $R_1$.  
Moreover, the grid cell boundaries are adjusted in accordance with their join partners in $R_1$. All updated cells $gc_j$ that have some overlap with any of the cells of $R_1$ together with the accumulated cardinality are used in the next join pair $(R_2, R_3)$ as the cells of $R_2$. This procedure is applied for all successive join pairs,   
in the end producing the query cardinality.

\subsection{Practical Challenges and Observations}
\label{ssec:practical_challengs_and_observations} 
When estimating the cardinality of range join queries, we distinguish between three cases. When the grid cells from one relation completely (do not) satisfy the join condition, the accuracy of the range join estimate depends only on the estimate of the individual table estimators. For instance, consider two attributes $c_1$ and $c_2$ belonging to the tables $R$ and $S$, respectively. For the join condition $c_1 < c_2$, when comparing two cells from each estimator, if we know the range of the dimensions $c_1$ and $c_2$ in both grid cells and we know that $c_{1max} < c_{2min}$, then the true cardinality is the multiplication of the cardinality of the individual cells, as every point in $c_1$ is definitely smaller than every point in $c_2$. However, since we rely on per-cell estimates of the individual estimators, even when the join condition is fully satisfied between the considered cells, \textit{a poor estimate of the estimators will contribute to worse estimation accuracy over range join queries.} 
The absolute error of this part is $\sum_{gc_{i} \in R_1}^{}{\sum_{gc_{j} \in R_2}^{}{|(card_{gc_i} * card_{gc_j}) - (est_{gc_i} * est_{gc_j})|}}$. Thus, a better accuracy in the estimator leads to a more accurate join estimation. 
For grid cells that partially satisfy the conditions, to produce representative samples, we need to \textit{capture the per-column data distribution}, if it is not present upfront. Although this can improve the accuracy, it increases the estimator size and the time required for training the models. 
From our evaluation, correctly capturing the distribution leads to a slight improvement over always assuming uniformity while drastically increasing the creation time.

Some existing approaches~\cite{NeuroCard} rely on precomputing the actual equi-join or its statistics to achieve accurate estimates.
Given the numerous combinations, it is \underline{not feasible} to materialize every possible range join with all potential expressions unless the join is predetermined before query estimation. Yet, optimizing an estimator for a single specific join variant is rarely practical.
Our algorithm avoids precomputing the join and can handle any range predicate and expression. While the dependence between tables is not captured through a join precomputation, partitioning the space permits accurate consideration or dismissal of large portions of the rows, contributing to accurate estimates.

\section{Experiments}
\label{sec:experiments}

We have realized our approach in Python. The CDF model has been implemented using SKlearn, and the autoregressive model using PyTorch. The experiments for the learned models, i.e., the supervised model for CDF estimation and the autoregressive model, are carried out on an NVidia GeForce RTX $2080$ Ti GPU. The computations of the grid and the competing non-learned approaches are executed on an Intel Xeon E$5$-$2603$ v$4$ CPU @$1.7$ GHz, $128$GB RAM.
 
The columns with textual values and dates are encoded using a dictionary prior to evaluation. For Grid-AR, the columns with continuous data are directly used to create the grid. 
To make the competing autoregressive model handle floating values, such columns are encoded using a dictionary before the evaluation.
When comparing the autoregressive model used in Grid-AR to the standalone autoregressive model, we use three \textit{layers} with $512$ neurons and train the model for ten \textit{epochs}. We use an \textit{embedding size} of $32$ for the input and output.  
We use per-column \textit{compression}~\cite{LMKG} for columns with more than $2000$ unique values by splitting the column into two subcolumns.  
The per-column CDF model is a Decision Tree Regressor where we varied the depth between $3$ and $6$.  

For single table queries,  
we report on the \textbf{accuracy}, \textbf{memory}, \textbf{training time}, and \textbf{estimation time}. For range joins, we report on the \textit{accuracy} and \textit{estimation time} using the same Grid-AR structures as for the single table estimates. 
For measuring the accuracy, we use $q\!-\!error = max(\frac{true(q)}{estimate(q)}, \frac{estimate(q)}{true(q)})$.

\subsection{Datasets}
\label{ssec:datasets} 
We evaluated the approaches on one synthetic and two real-world datasets. 
The characteristics of the datasets are detailed in Table~\ref{table:dataset-details}. 
\noindent{\underline{\textbf{TPC-H:}}} 
We use the \textbf{Customer} table from the \textit{TPC-H} synthetic benchmark with scaling factor one.  
The table has only one float attribute, which for standalone autoregressive models is dictionary mapped.  
\noindent{\underline{\textbf{Flight:}}}  
The real-world dataset \textbf{Flight} has flight sensor data from Germany, with  
nine attributes, where 
all numerical attributes are floating type, needing a dictionary mapping for the autoregressive model. 
\noindent{\underline{\textbf{Payment:}}} 
This real-world dataset contains information about the billing of a mid-sized company. We have selected the \textbf{Payment} relation with information about payments of bills issued to customers, such as the amount, date and type of payment,  
and others. Of the seven attributes, two have floating point values and need additional dictionary mapping. 

For \textbf{single table queries}, we generated $300$ queries for \textit{Customer} and $200$ queries for \textit{Flight} and \textit{Payment} with \textbf{varying numbers of query predicates} where the maximum number of predicates per dataset is specified in Table~\ref{table:dataset-details}. The predicates have one of the operations: $=, >, <, \le, \ge$. 
For \textbf{range join queries}, we used the tables \textbf{Customer} and \textbf{Flight} and created range join queries by performing self-joins for different join types and number of joins. For every table, we created $150$ range join queries with a varying number of equality and range predicates as for the single table queries. For both datasets, a \textit{multi-join query can include up to five tables}. 
Between two tables, for \textbf{Customer}, the queries have up to three join conditions and, for \textbf{Flight}, up to five. 
We group the range join types into \textbf{inequality}, i.e., inequality comparison of two attributes, and \textbf{range}, i.e., a point in interval or interval overlap join. We do not consider group-by or nested queries. 

\begin{table}[!t]
  \caption{Datasets and query characteristics.}
  \centering
  \renewcommand{\arraystretch}{0.99}
\scalebox{0.9}{
  \begin{tabular}{lrrr}
    \hline

{\textbf{Datasets}} & {\textbf{Customer}} & {\textbf{Flight} } & {\textbf{Payment}} \\
    \midrule
   {\textbf{Number of Rows}}  & \textbf{$n=150k$} & \textbf{$n=2.1M$} & \textbf{$n=8.8M$} \\
    \midrule
    Number of Cols. & $8$ & $9$ & $7$ \\
    Textual / Numerical Cols. & $5$ / $3$ & $3$ / $6$ & $3$ / $4$ \\
    Max Query Predicates & $5$ & $7$ & $5$ \\ 
    Max Join Tables & $5$ & $5$ & / \\
    \midrule
  \end{tabular}
}
\label{table:dataset-details} 
\end{table}

\begin{table*}[!t] 
  \caption{Estimation accuracy expressed as q-error.}
  \centering
  \renewcommand{\arraystretch}{0.95}
\scalebox{.925}{
  \begin{tabular}{lrrrrrrrrrrrr}
    \hline

{\textbf{Datasets}} & \multicolumn{4}{c}{\textbf{Payment}} & \multicolumn{4}{c}{\textbf{Flight}} & \multicolumn{4}{c}{\textbf{Customer}} \\
    \midrule
   {\textbf{Approach}}  & \textbf{Median} & \textbf{90th} & \textbf{Max} & \textbf{Average} & \textbf{Median} & \textbf{90th} & \textbf{Max} & \textbf{Average} & \textbf{Median} & \textbf{90th} & \textbf{Max} & \textbf{Average} \\
    \cmidrule(rl){2-5}
    \cmidrule(rl){6-9}
    \cmidrule(l){10-13} 
    EPostgres & $2.5$ & $21.7$ & $1.7*10^5$ & $3075.5$ & $2.4$ & $7.5$ & $115$ & $5.2$ & $2.4$ & $2.5$ & $3.5$ & $2.4$ \\
    Naru & $n/a$ & $n/a$ & $n/a$ & $n/a$ & $1.5$ & $9.5$ & $179.5$ & $7.6$ & $\textbf{1.07}$ & $1.4$ & $\textbf{2.7}$ & $1.2$\\
    CNaru &  $1.2$ & $\textbf{4}$ & $3*10^4$ & $540.8$ & $1.5$ & $10$ & $298$ & $11.1$ & $\textbf{1.04}$ & $\textbf{1.2}$ & $5.2$ & $\textbf{1.1}$ \\
    IAM & $\textbf{1.1}$ & $13.7$ & $6308.5$ & $116.3$ & $1.3$ & $\textbf{6}$ & $59.5$ & $\textbf{3.1}$ & $\textbf{1.09}$ & $1.6$ & $1366$ & $12.5$ \\
    \textbf{Grid-AR} & $1.38$ & $32$ & $\textbf{496}$ & $\textbf{20.1}$ & $\textbf{1.25}$ & $7$ & $\textbf{59}$ & $3.6$ & $\textbf{1.09}$ & $1.4$ & $7.8$ & $1.2$ \\
    \midrule
  \end{tabular}
}
\label{table:estimation_accuracy} 
\end{table*}

\subsection{Competitors}
\label{ssec:competitors}

By default, \textbf{Grid-AR} uses a CDF--based grid. 
We study the impact of uniform vs. CDF--based grid and grid resolution later-on separately.

\noindent{\underline{\textbf{Queries over single tables}}}: Since our approach aims to improve the autoregressive models by overcoming their problems when working with numerical and textual attributes, as our main competitor, we use the \textit{autoregressive model} architecture named Naru~\cite{Naru}. We use the code provided by the authors which includes progressive sampling to estimate queries with range predicates.  
Since highly heterogeneous attributes directly affect the model size and estimation time, we pair the model with a per-column compression technique~\cite{LMKG}. Thus, we consider two versions of Naru as competitors: one without compression, \textbf{Naru}, and the other with compression, \textbf{CNaru}.
For \textit{Payment}, we could not fit \textit{Naru} on our GPU even after  drastically reducing the batch and embedding size. Thus, we do not present results for the estimation time and accuracy for this dataset. We also consider \textbf{IAM}~\cite{DBLP:conf/edbt/MengWCZ022}, which is a modification on top of Neurocard~\cite{NeuroCard} that integrates Gaussian mixture models (GMM) with deep autoregressive models to separate the handling of continuous and categorical attributes. We use the publicly available implementation with the same parameter settings for the number of layers ($4$) and units ($256$, $128$, $128$, $256$), training epochs ($10$), number of GMM components for each attribute ($30$), samples for training ($10^6$), progressive sampling, and GMM, and number of subcolumns ($2$) as proposed by the authors. 
To showcase the memory consumption, we use the model with the smallest embedding size. 
To outline the effects of our approach, as a baseline for producing exact answers, we include PostgreSQL $14$ (\textbf{Postgres}). We also include the results of the classical assumption techniques paired with per-column 1D histograms of PostgreSQL (\textbf{EPostgres}).

\noindent{\underline{\textbf{Range join queries}}}: To the best of our knowledge, this is the first approach addressing cardinality estimation over range joins. As we do not have direct learned competitors that perform estimation, we compare to the state-of-the-art solutions that produce exact answers, namely, \textbf{DuckDB} v.0.9.2 and  \textbf{Tableau Hyper} v.0.0.13129, both used as Python packages. Both solutions offer faster execution, drastically \textbf{outperforming} PostgreSQL algorithms when handling range join queries. In addition to the exact competitors, we compare to the estimation approach included in PostgreSQL $14$ (\textbf{EPostgres}).

\subsection{Cardinality Estimation over Single Tables}
\noindent{\underline{\textbf{Accuracy}}:  
The accuracy results (Table~\ref{table:estimation_accuracy}) are an average of \textit{five} executions.  
We show the average, median, $90$th percentile, and maximal q-error.  
A higher q-error means worse estimation accuracy.  

For all datasets, it is evident that Grid-AR produces comparable results or slightly less accurate when compared to the best approach, IAM, in terms of median accuracy. However, for \textit{Payment}, Grid-AR drastically outperforms the competitors when considering the average q-error. For the \textit{Flight} dataset, Grid-AR produces estimates with the best median q-error while having a comparable  average q-error to IAM. For the higher percentiles, CNaru has worse estimation accuracy than Naru. The reason for this is the introduced compression, which affects the correlation between the columns, introducing more complex interconnections for the model to learn. This also applies to Grid-AR since we always use the compressed version of the autoregressive model. 
EPostgres produces the worst estimation accuracy, visible for highly heterogeneous datasets such as \textit{Payment}, producing substandard results that cannot be considered in any real-world scenario. 

\begin{table}[!t]
  \caption{Training time in seconds represented as the average time for training per epoch.}
  \centering
  \renewcommand{\arraystretch}{0.95}
\scalebox{0.95}{
  \begin{tabular}{lrrr}
    \hline

{\textbf{Datasets}} & {\textbf{Payment}} & {\textbf{Flight}} & {\textbf{Customer}} \\
    \midrule
   {\textbf{Approach}}  & \textbf{Average Time} & \textbf{Average Time}  & \textbf{Average Time} \\
    \cmidrule{2-4}
    Naru & n/a & $2600$ & $39.1$ \\
    CNaru & $136.1$& $39.2$ & $4.1$ \\
    IAM & $270.8$& $744.6$ & $36.1$ \\
    \textbf{Grid-AR} & $\textbf{85.5}$ & $\textbf{24}$ & $\textbf{3.2}$\\
    \midrule
  \end{tabular}
}
\label{table:training_times}
\end{table}

\begin{table}[!t]
  \caption{Estimation time in milliseconds represented as the median and average time per query.}
  \centering
  \renewcommand{\arraystretch}{0.95}
\scalebox{0.8}{
  \begin{tabular}{lrrrrrr}
    \hline

{\textbf{Datasets}} & \multicolumn{2}{c}{\textbf{Payment}} & \multicolumn{2}{c}{\textbf{Flight}} & \multicolumn{2}{c}{\textbf{Customer}} \\
    \midrule
   {\textbf{Approach}}  & \textbf{Average} & \textbf{Median} & \textbf{Average} & \textbf{Median} & \textbf{Average} & \textbf{Median} \\
    \cmidrule{2-7}
    Postgres & $553.4$ & $539$ & $232.1$ & $223.8$ &  $20.2$ & $19.3$\\
    \midrule[0.5pt]
  EPostgres & $\textbf{0.37}$ & $\textbf{0.29}$ & $\textbf{0.33}$ & $\textbf{0.31}$ & $\textbf{0.2}$ & $\textbf{0.18}$\\
    Naru & $n/a$ & $n/a$ & $509.6$ & $527.6$ & $62.6$ & $82.1$ \\
    CNaru & $17$ & $8.5$ & $459.1$ & $489.8$ & $55.2$ & $72.9$\\
    IAM & $15.9$ & $12.7$ & $15.6$ & $15.2$ & $19$ & $18.7$\\
    \textbf{Grid-AR} & \colorbox{gray!20}{$14.4$} & \colorbox{gray!20}{$7.4$} & \colorbox{gray!20}{$8.7$} & \colorbox{gray!20}{$6.8$} & \colorbox{gray!20}{$1.5$} & \colorbox{gray!20}{$1.5$} \\ 
    \midrule
  \end{tabular}
}
\label{table:estimation_times} 
\end{table}

\begin{figure*}[!t] 
\centering
\scalebox{0.85}{
\subfloat[Payment]{
    \centering
    \includegraphics[width=0.33\linewidth]{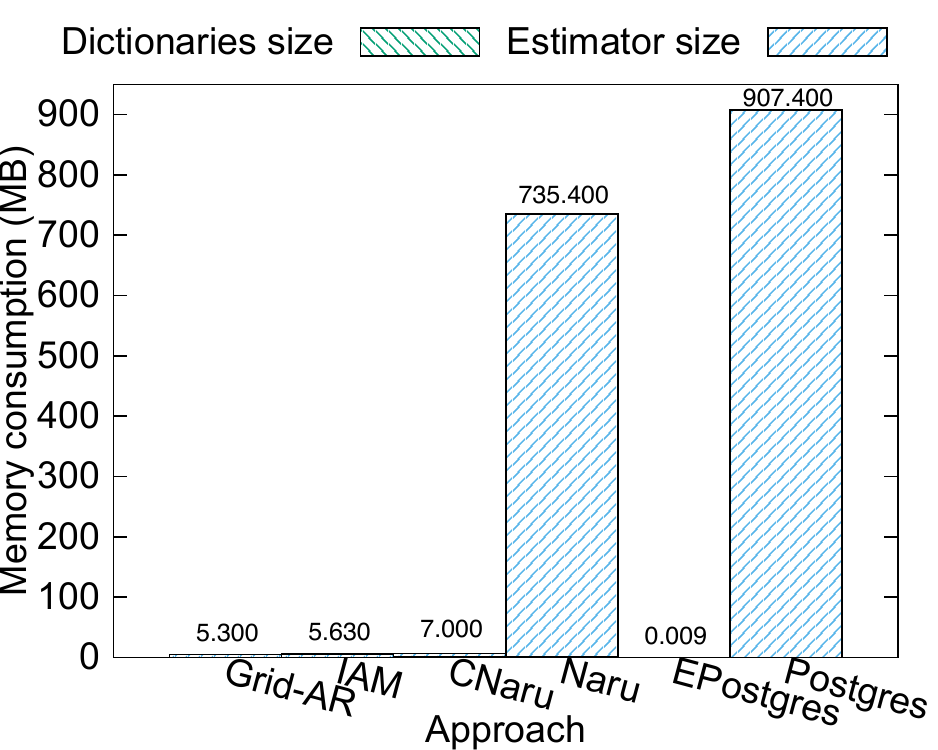}
    \label{fig:memory_rw}
}%
\subfloat[Flight]{
    \centering
    \includegraphics[width=0.33\linewidth]{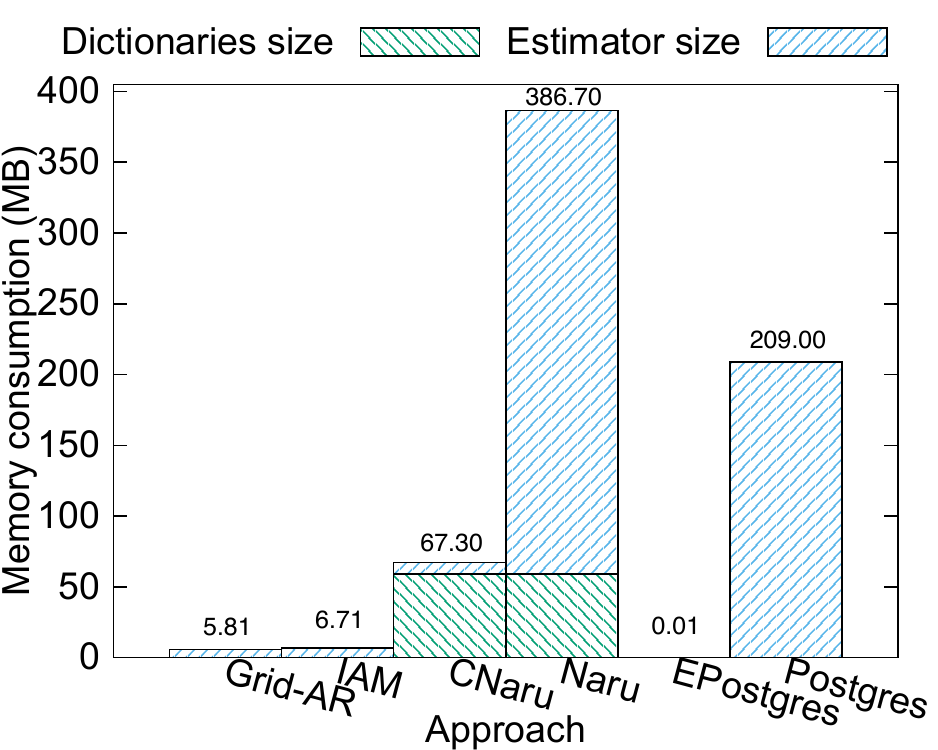}
    \label{fig:memory_flight}
}%
\subfloat[Customer]{
    \centering
    \includegraphics[width=0.33\linewidth]{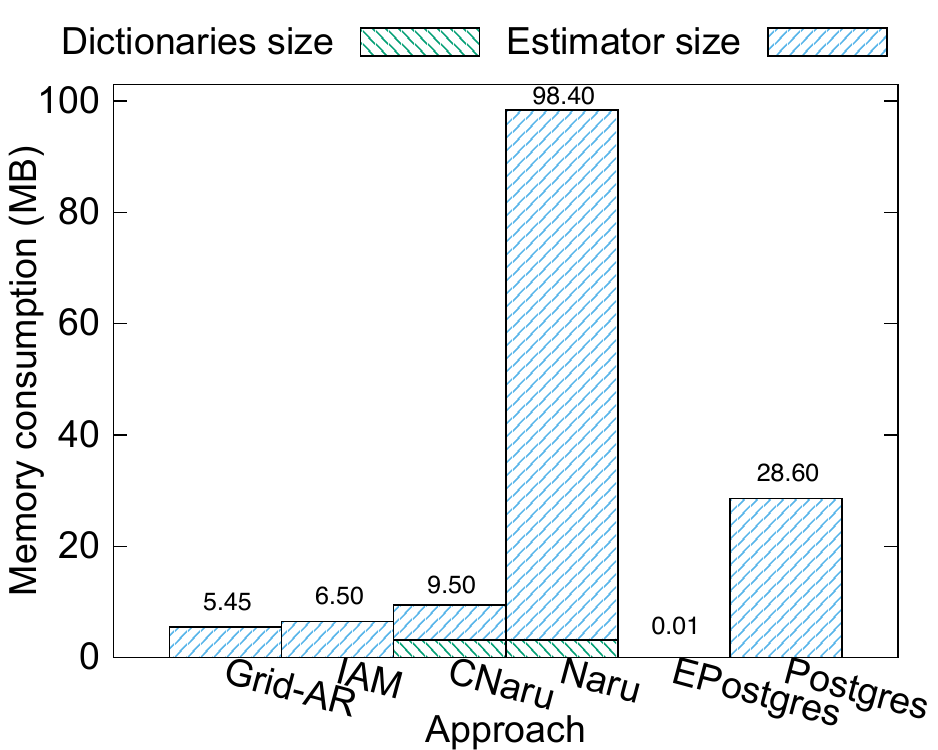}
    \label{fig:memory_tpch}
}
}
\caption{Memory consumption in megabytes considering the size of both the estimator and the dictionary mapping.}
\label{fig:memory_consumption_exp} 
\end{figure*}

\noindent{\underline{\textbf{Training and estimation time}}}: 
Next, we detail the training time for the learned approaches  
and the estimation time for all approaches for the three datasets. When considering \textit{Naru} and \textit{CNaru}, for the progressive sampling-based estimation, we always select $1000$ samples.  
All autoregressive models have our standard setting. 

The training time results (Table~\ref{table:training_times}) underpin the importance of compression for highly heterogeneous datasets. 
Going a step beyond, our Grid-AR approach groups points into buckets, yielding a smaller volume of training data. Grid-AR requires \textit{less training time} while managing to produce estimates \textit{faster}. 
For all considered datasets, Grid-AR consistently achieves the best training time.

When analyzing the estimation times presented in Table~\ref{table:estimation_times}, it is clear that Grid-AR \textit{outperforms all the considered approaches except for EPostgres}, where in most scenarios, it produces results that are orders of magnitude better than those of the competitors.  
The estimation time of Grid-AR, for some datasets, even contends the estimation time of the per-column 1D histograms that are used by EPostgres.  
For the \textit{Customer} and \textit{Flight} datasets, the difference in the estimation time to the competitors is especially prominent. The main reason for this is the presence of more numerical attributes. For Grid-AR, these attributes will be included in the grid, whereas for Naru and CNaru, proper value mapping and optional compression will be performed.  
Due to the large number of numerical columns, the competitors have an increased number of attributes in the model, directly influencing the iterative sampling required for the range queries,  
plus the additional overhead of finding the integer mapping. 
Although the input will be smaller for IAM, the iterative sampling algorithm will still need to be performed.  
Grid-AR avoids the costly operations of mapping and iterative sampling required by the competitors.  
Producing estimates fast is crucial for any estimation approach, so these results further show the importance of our solution of splitting the estimation to consider numerical and text-valued columns independently.

\noindent{\underline{\textbf{Memory consumption}}}:   
When measuring the memory, we consider two parts: the memory of the estimator and the memory of the value-integer mappings. For Grid-AR, we store the grid structure, autoregressive model, and the mappings for the textual columns. For the autoregressive models, we store the mappings of the textual and floating-point columns in addition to the model. For Postgres, we consider table size, and for EPostgres, the sum of the sizes of the per-column 1D histograms. 
The required memory, in megabytes, is shown in \figurename~\ref{fig:memory_consumption_exp}.  
After EPostgres, Grid-AR occupies the least memory for all datasets. This difference is especially noticeable for the \textit{Flight} dataset (\figurename~\ref{fig:memory_flight}) due to the number of columns that have floating-point values. Since more than half the columns are of this type, Naru and CNaru need to store additional mapping for them, as shown by the \textit{Dictionary Size} value.  
For the \textit{Payment} dataset (\figurename~\ref{fig:memory_rw}), since there is only one column that has floating-point values, the difference between the required memory of Grid-AR and CNaru is smaller.  
For IAM, the usage of Gaussian mixture models requires slightly more memory than our Grid-AR approach. 

Grid-AR has the best balance among all approaches by requiring the least memory and producing results with comparable accuracy to the most accurate approach but in a drastically faster time. 

\begin{table}[!t] 
  \caption{Accuracy (q-error) of estimator variants when varying the total number of grid cells (Payment).}
  \centering
  \renewcommand{\arraystretch}{0.99}
\scalebox{0.73}{
  \begin{tabular}{lrrrrrrrr}
    \hline
{\textbf{Grid cells}} & \multicolumn{2}{c}{\textbf{10 000}} & \multicolumn{2}{c}{\textbf{20 000}}& \multicolumn{2}{c}{\textbf{30 000}} & \multicolumn{2}{c}{\textbf{50 000}} \\
    \midrule
   {\textbf{Approach}}  & \textbf{Median} & \textbf{Avg} & \textbf{Median} & \textbf{Avg} & \textbf{Median} & \textbf{Avg} & \textbf{Median} & \textbf{Avg}\\
    \cmidrule(rl){2-3}
    \cmidrule(rl){4-5}
    \cmidrule(l){6-7}
    \cmidrule(l){8-9}
    Uniform& $1.2$ & $459$ & $1.1$ & $417$ & $1.1$ & $409$ & $1.1$ & $375$ \\
    CDF & $1.4$ & $84$ & $2$ & $22$ & $1.3$ & $20$ & $1.6$ & $14$\\
    \midrule
  \end{tabular}
}
\label{table:grid_accuracy_varying_cells} 
\end{table}

\subsection{Grid-AR Analysis}
\label{ssec:exp_model_analysis}
 
To showcase the importance of correctly capturing the per column distributions, 
we analyze the effects of how we split the ranges per dimension in the grid of Grid-AR.  
The version that uses uniformity assumption is called \textit{Grid-AR Uniform}, and the version using per dimension CDFs for creating the cells is called \textit{Grid-AR CDF}. We compare the versions by varying the number of grid cells.  

\noindent{\underline{\textbf{Accuracy}}}:  
When analyzing the results in Table~\ref{table:grid_accuracy_varying_cells}, we notice that as the size of the grid increases, as expected, we observe an improvement in the estimation accuracy, 
especially prominent in the average accuracy. 
However, the median accuracy, even for the smallest grid, is already of acceptable quality and comparable to the largest grid structure. Grid-AR CDF has better average accuracy for all considered grid sizes, where starting from the smallest grid ($10000$ cells), it drastically outperforms Grid-AR Uniform.

\begin{figure}[!t]
\vspace{-3mm}
\centering 
\subfloat[Estimation time]{
    \centering
    \includegraphics[width=0.49\linewidth]{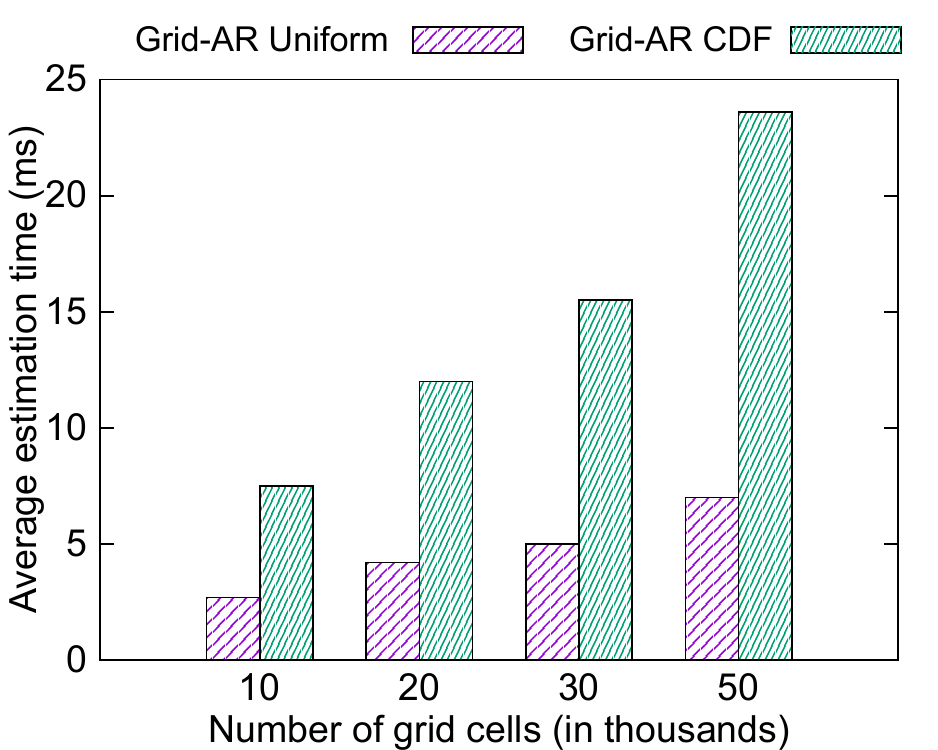}
    \label{fig:estimation_time_grid_varying_cells}
}%
\subfloat[Memory consumption]{
    \centering
    \includegraphics[width=0.49\linewidth]{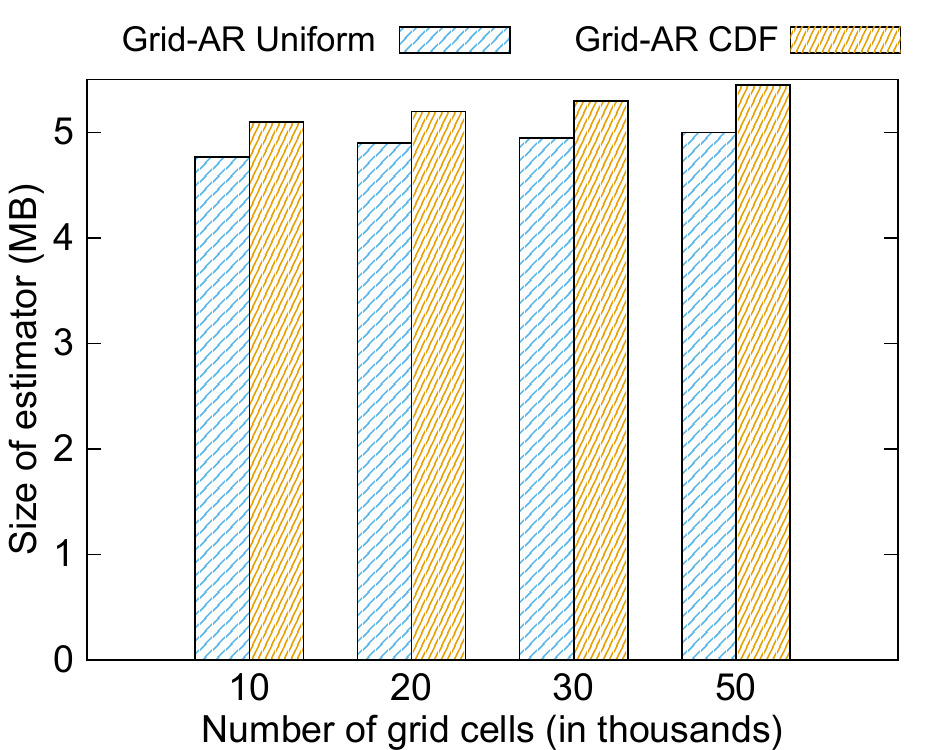}
    \label{fig:memory_grid_varying_cells}
}
\caption{Estimation time in milliseconds (left) and memory consumption in megabytes (right) when varying the total number of grid cells (Payment).}
\label{fig:memory_estimation_grid_varying_cells} 
\end{figure}

\noindent{\underline{\textbf{Estimation time}}}:   
We depict the median estimation time in \figurename~\ref{fig:estimation_time_grid_varying_cells}. 
The increase in the number of cells of the grid structure directly contributes to increased estimation time (\figurename~\ref{fig:estimation_time_grid_varying_cells}). However, although there is a sharp increase in the number of cells, the estimation time is still in the range of milliseconds. 
In contrast to Grid-AR Uniform, Grid-AR CDF requires additional computation of the CDF to gather the qualifying cells for a user-imposed query. However, the additional overhead is not significant, especially when considering the improvement in accuracy (Table~\ref{table:grid_accuracy_varying_cells}).

\noindent{\underline{\textbf{Memory consumption}}}:   
For both versions, as the number of cells increases, there is almost no difference in the memory consumption of the grid structures (\figurename~\ref{fig:memory_grid_varying_cells}). The constituent autoregressive model occupies the largest portion of the memory for both versions. 
Despite the increase in the number of cells for each version, this difference, although noticeable, will not be highly pronounced due to our use of per-element compression for the remaining columns. In addition, the grid stores the bucket boundaries for the numerical attributes of the relation. Thus, as the number of cells increases, so does the number of bucket boundaries for the attributes, resulting in increased information that needs to be stored.  
Grid-AR CDF has slightly larger memory than Grid-AR Uniform due to the additional overhead from storing a regressor per dimension. 

\begin{table}[!t]
  \caption{Accuracy (q-error) for range join queries over two tables for different join types on Customer and Flight.}
  \centering
  \renewcommand{\arraystretch}{1}
\scalebox{0.85}{
  \begin{tabular}{llrrrr}
    \hline
\textbf{Approach} & \textbf{Query Type} & \multicolumn{2}{c}{\textbf{Inequality} }& \multicolumn{2}{c}{\textbf{Range}} \\
 \midrule 
  & {\textbf{Dataset}} & \textbf{Median} & \textbf{Average} & \textbf{Median} & \textbf{Average} \\
    
    \cmidrule(rl){2-2}
    \cmidrule(rl){3-4}
    \cmidrule(l){5-6}
    EPostgres & Flight & $10$ & $169.5$ & $21.8$ & $303.1$\\
    & Customer & $3.85$ & $7$ & $5.1$ & $8.5$ \\
     \midrule[0.5pt]
     \textbf{Grid-AR} & Flight & $\textbf{1.44}$ & $\textbf{5.9}$ & $\textbf{1.56}$ & $\textbf{11.2}$\\
     & Customer & $\textbf{1.25}$ & $\textbf{4.1}$ & $\textbf{1.2}$ & $\textbf{1.8}$ \\
     
    \midrule

  \end{tabular}
}
\label{table:grid_accuracy_varying_query_type_range} 
\end{table}

\begin{figure}[!t]
\vspace{-4mm}
\centering 
\subfloat[Customer]{
    \centering
    \includegraphics[width=.48\linewidth]{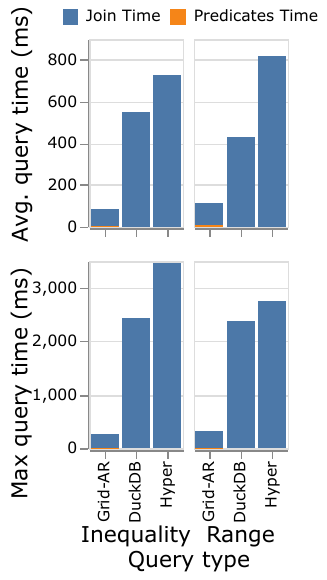}
    \label{fig:estimation_time_range_cust}
}%
\subfloat[Flight]{
    \centering
    \includegraphics[width=.47\linewidth]{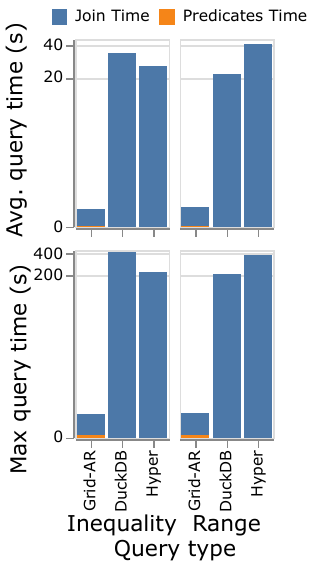}
    \label{fig:estimation_time_range_flight.pdf}
}
\caption{Average and maximal estimation time per query type for range join queries over two tables on \textbf{Customer} (left) and \textbf{Flight} (right). For Flight, y-axis is log scaled.}
\label{fig:estimation_time_range_full} 
\end{figure}

\subsection{Cardinality Estimation for Range Joins}
\label{ssec:exp_card_estimation_range_joins} 
We compare our approach for cardinality estimation over range joins to the considered competitors. We report the \textit{estimation time} of the approaches and the \textit{accuracy} of our approach and EPostgres. We use the same Grid-AR per dataset as for the single table query evaluation. Since we evaluate range joins with a self-join,  
there is one Grid-AR for every table in the range join query. The memory of our approach, per table, remains the same as for the previous evaluation. The competitors do not store additional indexes.

\noindent{\underline{\textbf{Accuracy}}}:  
The estimation accuracy shown through the q-error grouped by the different query types is depicted in Table~\ref{table:grid_accuracy_varying_query_type_range}. From the $150$ queries for every dataset, the number of inequality and range join queries is proportional. From the results, it is evident that Grid-AR \textit{drastically outperforms} EPostgres. For Flight,  
the inequality joins achieve better overall accuracy than the range joins. This is because in range join queries, comparing cells from the estimators of the different tables leads to more cells that partially overlap with one another. This, in turn, triggers the computation of the overlap estimation of the cells, further affecting the per-cell estimated cardinalities. Note that the estimates for the range join queries (inequality and range) directly depend on the quality of the individual Grid-AR structures. If the underlying structure produces inferior estimates for the predicates over the individual tables, their accuracy further deteriorates when capturing the range join.

\noindent{\underline{\textbf{Estimation time}}}:  
We present the average and maximal estimation time per query type in \figurename~\ref{fig:estimation_time_range_full}.  
For Grid-AR, we split the time into \textit{predicates time}, which is the time that the per-table estimators require for estimating the qualifying cells for the query and their cardinality, and \textit{join time}, which is the time for computing the join result from the qualifying cells. 
From the depicted results, it is evident that our approach achieves the best estimation time for both datasets. When considering the \textit{Flight} dataset (\figurename~\ref{fig:estimation_time_range_flight.pdf}, note the log scale) whose table is significantly larger than the \textit{Customer} table, the Grid-AR--based approach \textit{produces estimates in milliseconds} where the competitors require, on average, more than $20$ seconds.  
When looking at the maximal query time, it is evident that Grid-AR can perform efficiently even for not-so-selective queries. For EPostgres, the estimation time is between $0.5$ and $1$ ms, irrespective of the query type or cardinality.

\begin{table}[!t] 
  \caption{Median accuracy (q-error) of Grid-AR for three, four, and five tables for the range join queries for different join types on Customer and Flight.}
  \centering
  \renewcommand{\arraystretch}{1}
\scalebox{0.71}{
  \begin{tabular}{llrrrrrrrrr}
    \hline
\textbf{Approach} & \textbf{Query Type} & \multicolumn{3}{c}{\textbf{Inequality} }& \multicolumn{3}{c}{\textbf{Range}} & \multicolumn{3}{c}{\textbf{Combination}} \\
 \midrule 
  & {\textbf{Dataset}} & \textbf{3} & \textbf{4} & \textbf{5} & \textbf{3} & \textbf{4} & \textbf{5} & \textbf{3} & \textbf{4} & \textbf{5} \\
    
    \cmidrule(rl){2-2}
    \cmidrule(rl){3-5}
    \cmidrule(l){6-8}
    \cmidrule(l){9-11}
    EPostgres & Flight & $10.5$ & $147.9$ & $71.1$ & $15.2$ & $264.7$ & $16.3$ & $9.1$ & $36.4$ & $153$\\
    & Customer & $2.4$ & $5.9$ & $4.9$ & $14.1$ & $63.9$ & $29.9$ & $5.8$ & $4$ & $6$ \\
     \midrule[0.5pt]
     \textbf{Grid-AR} & Flight & $7.8$ & $7.8$ & $29.5$ & $9$ & $24$ & $11.9$ & $3.9$ & $13$ & $45$ \\
     & Customer & $1.5$ & $1.8$ & $1.7$ & $1.8$ & $2.4$ & $2.3$& $1.6$ & $1.7$ & $1.8$ \\
     
    \midrule
  \end{tabular}
}
\label{table:grid_multi_table_range_joins} 
\end{table}

\noindent{\underline{\textbf{Multi-table range joins}}}: To showcase the generalizability of our range join approach, we created join queries that involve three, four, and five tables, consisting of only inequality joins, only range joins, and a combination of both. The accuracy results  
are shown in Table~\ref{table:grid_multi_table_range_joins}.  
Grid-AR continues to outperform EPostgres in terms of estimation accuracy for the different join queries and number of involved tables. For the uniformly distributed \textit{Customer} dataset, Grid-AR produces estimates that are almost always at least twice as good as those of EPostgres.  
For \textit{Flight}, our approach can potentially benefit from capturing the per-column distribution instead of always assuming uniformity when sampling for range join estimation. 
EPostgres took the same average time as for the queries with two tables. For all join types and number of tables, Grid-AR took on average $600$ milliseconds for \textit{Flight} and $21$ milliseconds for \textit{Customer}. For DuckDB, the average time was $11$ seconds for \textit{Flight} and $116$ seconds for \textit{Customer}. For Hyper, the average time was $32$ seconds for \textit{Flight} and $71$ seconds for \textit{Customer}.  

\subsection{Overall Performance}
\label{ssec:overall_performance}

To examine the effects of Grid-AR  
with concerning the end-to-end query execution, we have integrated our approach into PostgreSQL $14$ following the procedure from related work~\cite{DBLP:conf/edbt/MengWCZ022,DBLP:conf/sigmod/CaiBS19}. We ingest the cardinality estimates in the query optimizer of Postgres such that for every query we incorporate the estimates of the subqueries. The modified Postgres then uses the estimates to generate a query plan for which we measure the end-to-end query execution time. Since none of  the competing estimation approaches support range joins, to compare to them, we created $100$ queries for the Flight dataset consisting of equality joins from the same single table queries from the comparison to the competitors, having up to five tables that are being joined. The results are depicted in Table~\ref{table:end_to_end_time_postgres}. In addition, to evaluate the range joins, we used $300$ queries that consist of range joins, inequality joins, or a combination of both, having up to five joining tables, and compared Grid-AR to the native Postgres estimation and having the exact cardinalities as estimates. The results are shown in Table~\ref{table:end_to_end_time_range_postgres}. We measure the end-to-end query execution time, planning time, and the improvement (percentage) over the native Postgres estimation. The improvement is calculated as (time of EPostgres - approach time) / time of EPostgres.

\begin{table}[!t] 
  \caption{End-to-end time for single table queries (Flight).}
  \centering
  \renewcommand{\arraystretch}{1}
\scalebox{0.83}{
  \begin{tabular}{lrrrr}
    \hline
  \textbf{Approach} &  \multicolumn{2}{c}{\textbf{End-to-end Time}}  & \multicolumn{2}{c}{\textbf{Improvement}} \\
 \midrule
 &  \textbf{Median} & \textbf{Average} & \textbf{Median} & \textbf{Average} \\
    \midrule
    EPostgres & $0.5$s & $3.2$s &  -- & -- \\
    Exact Card. (optimal) & $0.48$s & $2.93$s & $4\%$ & $8.43\%$ \\
    Naru & $0.49$s & $4.82$s & $2\%$ & -$50.62\%$  \\
    CNaru & $0.49$s & $3.2$s & $2\%$ & $0\%$  \\ 
    IAM & $0.5$s & $3.2$s & $0\%$ & $0\%$   \\
    \textbf{Grid-AR} & $0.49$s & $3.1$s & $2\%$ &  $3.12\%$  \\
     
    \midrule
  \end{tabular}
}
\label{table:end_to_end_time_postgres} 
\end{table}
\begin{table}[!t]
  \caption{End-to-end time for range join queries (Flight).}
  \centering
  \renewcommand{\arraystretch}{1}
\scalebox{0.83}{
  \begin{tabular}{lrrrr}
    \hline
 \textbf{Approach} &  \multicolumn{2}{c}{\textbf{End-to-end Time}}  & \multicolumn{2}{c}{\textbf{Improvement}} \\
 \midrule
 &  \textbf{Median} & \textbf{Average} & \textbf{Median} & \textbf{Average} \\
    \cmidrule(rl){2-3}
    \cmidrule(rl){4-5}
    EPostgres & $2$s & $9.35$s & -- & --\\
    Exact Card. (optimal) & $1.43$s & $8.23$s & $28.5\%$ & $11.98\%$\\
    \textbf{Grid-AR} & $1.44$s & $8.28$s & $28\%$ & $11.44\%$ \\
     
    \midrule
  \end{tabular}
}
\label{table:end_to_end_time_range_postgres}
\end{table}

For single table estimates (Table~\ref{table:end_to_end_time_postgres}), 
Grid-AR achieves end-to-end query runtime that is close to the optimal performance. However, the improvements in end-to-end runtime are not proportional to the improvements in the quality of the estimates. This is because the query optimizer can still generate the same plan even if the estimated cardinalities are different. Moreover, even though the estimates for the subqueries differ significantly for some queries, they can still be ranked in the same order, resulting in the same query plan. 
For range join queries (Table~\ref{table:end_to_end_time_range_postgres}), the improvements in the query runtime are more prominent, especially for the median query execution time. It is visible that Grid-AR has almost the same performance as the optimal approach with the exact query cardinalities. Our empirical evaluation demonstrates the importance of accurate cardinality estimates, especially when considering the unexplored field of range joins. 


\section{Related Work}
\label{sec:related_work}

\noindent{\underline{\textbf{Learned cardinality estimation:}}} 
Approaches based on machine learning models try to improve the cardinality estimates in different ways, e.g., by relying on prior query monitoring~\cite{LEO}, multi-set convolutional networks~\cite{MSCN}, 
simple deep learning models~\cite{SimpleModels}, pure data-driven models~\cite{DeepDB}, local-oriented approaches ~\cite{LocalDeepLearningModels}, containment rates~\cite{LearningContainmentRates} and other neural network architectures for relational operators~\cite{LiuXYCZ15} or multi-dimensional range predicates~\cite{DuttWNKNC19} .  
Others~\cite{DQ, REJOIN1, REJOIN2, Neo, StateRepresentationRL} use reinforcement learning for optimal plan generation 
or learning for approximate query processing~\cite{TupleBubbles, AQPGM, AQPDGM, DBEst, LAQP}. 
Naru~\cite{Naru} achieves accurate cardinality estimates with autoregressive models and a Monte Carlo integration technique.  
NeuroCard~\cite{NeuroCard} builds upon Naru by introducing per-column compression to reduce the memory and learns over samples from the join to enable efficient join cardinality estimation. Hasan et al.~\cite{MultiAttributeCE} suggest both autoregressive and supervised models for accurate estimates. LMKG~\cite{LMKG} uses autoregressive models with per-column compression and supervised models to estimate cardinalities over knowledge graphs. For the comparison, we utilize their compression for the single table estimators. Flat~\cite{DBLP:journals/pvldb/ZhuWHZPQZC21} uses factorized sum split product networks for accurate and fast cardinality estimation. Their results further underpin the accuracy of deep autoregressive models and the need for faster estimation.   
BayesCard~\cite{BayesCard,DBLP:journals/pvldb/HanWWZYTZCQPQZL21} creates an ensemble of Bayesian networks to model the table's distribution,  
and employs progressive sampling and variable elimination inference algorithms for  
multi-table join queries. 
Differently Flow-Loss~\cite{FlowLoss} is a novel loss function for training cardinality models that focuses on optimizing estimates critical for query optimization,   
resulting in improved runtimes despite potentially worse estimation accuracy.
Meng et al.~\cite{DBLP:conf/edbt/MengWCZ022} combine Gaussian mixture models and autoregressive models to improve estimation for continuous attributes. Their model, though similar to ours, demands considerably more training time and complexity, and does not consider range joins. 
Tuple Bubbles~\cite{TupleBubbles} is an AQP approach for single-table and join queries that groups tuples into bubbles, represented through unsupervised estimators.
FactorJoin~\cite{FactorJoin} formulates the estimation of join cardinality as an inference involving only learned distributions over single tables.
\textit{Note that while specific algorithms permit join queries, they are tailored for equality joins rather than range joins.}

\noindent{\underline{\textbf{Range join optimization:}}}
For range joins, typical database implementations frequently rely on inefficient  
algorithms that compare every pair of rows.
However, recent research has concentrated on addressing this drawback. 
One of them is DuckDB~\cite{DuckDB}, a high-performance analytical database management system with a rich SQL dialect supporting advanced features, where one of them is range joins optimized with fast sorting algorithms. 
Hyper~\cite{Hyper} allows fast query processing on large and complex datasets, with the ability to execute range join queries much faster than PostgreSQL. Reif and Neumann~\cite{DBLP:journals/pvldb/ReifN22} develop a kd-tree-based algorithm for range joins, 
 that can be made parallel both during building and probing. They prove its superiority by integrating the algorithm into their database system Umbra~\cite{Umbra}. \textit{Note that all algorithms mentioned consider complete execution and do not discuss cardinality estimation.}
Our approach incorporates all the benefits of prior methods, utilizing ``bins'' via grid cells, quick sorting to streamline comparisons and skipping, and the possibility for parallel execution.

\section{Conclusion} 
We proposed Grid-AR, a novel cardinality estimator for single table and range join queries.
By combining a grid structure with autoregressive models, Grid-AR distinguishes between column types, separating the continuous from categorical ones. 
Grid-AR helps circumvent two intrinsic problems of autoregressive cardinality estimators. 
Through partitioning in the grid, we avoid storing large dictionary mappings of the numerical columns, greatly diminishing the space needed for the estimator.
To reduce the search space and bypass the slow iterative sampling, we proposed a faster batch algorithm for range queries that leverages the grid. 
Using Grid-AR, we developed an algorithm that estimates the cardinality over range join queries efficiently.
The experimental evaluation showed that Grid-AR exceeds the state-of-the-art approaches regarding memory and speed while achieving comparable accuracy when applied over single tables.
We further demonstrated the importance of a cardinality estimator for range join queries that delivers quality estimates drastically faster than the exact approaches.

\balance
\bibliographystyle{ACM-Reference-Format}
\bibliography{main}

\end{document}